\documentclass[12pt]{article}
\usepackage{amssymb,amsmath,amsthm}
\usepackage{rotating}
\usepackage{graphicx,comment}
\usepackage{color}

\excludecomment{omitext}

\begin{document}


\begin{center} 

{ \bf  A paradox in Hele-Shaw  displacements   } 

 Gelu Pa\c{s}a

\end{center}

{ \small
{\bf Abstract}. 

We study the Hele-Shaw immiscible displacements when all surfaces  tensions on the  
interfaces are zero.
The  Saffman-Taylor  instability  occurs when a less viscous fluid is displacing 
a more viscous one, in a rectangular Hele-Shaw cell. 
We prove that an intermediate liquid with a variable viscosity can almost suppress 
this instability.    
On the contrary,  a large number of constant viscosity liquid-layers inserted   
between the initial fluids  gives  us boundless  growth rates  with respect to the 
wave  numbers of perturbations.  The same amount of intermediate liquid is used in 
both cases.


\vspace{0.5cm}



\begin{center} 

{\bf   1. Introduction} 

\end{center}

We consider  a  Stokes flow in a Hele-Shaw cell (see \cite{HS}) parallel  
with the plane $xOy$. 
The  thickness of the gap between the cell plates is $b$. The gravity effects 
are neglected. 
The  viscosity,  velocity  and  pressure are denoted by  $\nu,  {\bf u}=
(u,v,w), p$. As $b$ is very small, we neglect $w$. The flow equations are 
\begin{equation}\label{HS-4}
 p_x = -  \frac{b^2}{12 \nu } <u>, \quad 
 p_y = -  \frac{b^2}{12 \nu } <v>, \quad p_z=0, 
\end{equation}
$$   u_x+v_y=0,                                                          $$
where the lower indices  $x,y,z$ are denoting the partial derivatives and
  $< F > =(1/b) \int_0^b F dz $.
The above equations are similar to the  Darcy's law for the flow in  a porous 
medium  with the  permeability  $(b^2/12) $ - see ~\cite{LAMB},   \cite{BE}.


A sharp interface exists between two immiscible  displacing fluids in a Hele-Shaw cell.  
This flow-model can be used 
to study the secondary oil-recovery process: the oil (with low pressure) contained in a 
porous medium is obtained  by pushing it with a second  
displacing fluid.  Saffman and Taylor   ~\cite{SAFF-TAY}  proved the well know  result: 
the interface is unstable  when the displacing fluid  is less viscous. Moreover,  the 
fingering   phenomenon appears in this case - see also  ~\cite{SAFF},  ~\cite{HOM}.
The  Saffman - Taylor  growth constant   is unbounded in terms of the wave numbers if
the surface tension on the interface is missing.
On the contrary, a  surface tension on the interface is limiting the range of disturbances
which are unstable - see the formula (11)  in \cite{SAFF-TAY}.

The optimization of  displacements in porous media were studied  in ~\cite{AL-HUS-1},   
~\cite{AL-HUS-2},  ~\cite{CHEN},  ~\cite{DIAZ},  ~\cite{SUDARYANTO}.

An intermediate fluid  with  a variable  viscosity in a middle layer between the 
displacing fluids   can minimize  the  Saffman-Taylor instability, when 
the surface  tensions acting  on  interfaces {\it are not zero} - see  the experimental 
and numerical results  given in \cite{GILJE}, ~\cite{GOR-HOM-1},  ~\cite{GOR-HOM-2},  
~\cite{SHAH},  ~\cite{SLOBOD},    ~\cite{UZOIGWE}.  
A linear stability  analysis of such {\it three-layer} Hele-Shaw flow was performed in   
~\cite{DAR-PA-1}, ~\cite{DAR-PA-2},  ~\cite{DAR-PA-3} and  exact formulas of the growth 
constants  were given,  for variable  and  constant  intermediate viscosities. Due to the 
surface tensions on the interfaces, the  obtained growth constant are bounded  in terms 
of the wave numbers.

The Hele-Shaw displacement  with  $N$  intermediate layers (the multi-layer Hele -
Shaw model) when all surface tensions   are {\it different from  zero} was studied 
in ~\cite{DAR-2},  ~\cite{DAR-DING},  \cite {D7} , \cite {D6}.  Only upper bounds 
of the growth rates were obtained in terms of the problem data. In the case of  
intermediate viscosities with  positive jumps in the flow direction, in ~\cite{DAR-2} 
was proved that the corresponding growth rates  tend to zero when the  number  of the 
intermediate  layers is very large   and the  surface tensions verify some  conditions.

In this paper we point out a paradox concerning the stability of Hele-Shaw displacements
without surface tensions on the interfaces. 
For this,  we study the following two ``scenarios''.

First, we consider  a large number of constant viscosity liquid-layers inserted in  
the intermediate region,  with positive viscosity  jumps  in the  flow direction. 
We get inferior limits for growth constants, unbounded as functions of the wave numbers.
Therefore   the  multi-layer Hele-Shaw model  studied  in  ~\cite{DAR-2},  ~\cite{DAR-DING},  
\cite {D7} , \cite {D6}   is useless when   all surface tensions on the interfaces  are zero - 
 the displacement is unstable.

In the second case, a   liquid with a continuous  linear increasing viscosity   is considered 
between the less viscous displacing  fluid and the displaced one. We obtain  an upper bound of 
the growth rate of perturbations, independent of the wave numbers. The flow is  almost stable  
if the intermediate region is is long enough. 

It is important to underline that  we use the same amount of intermediate liquid  in both cases.

The paper is laid as follows. 
In section 2   we describe the three-layer Hele-Shaw model introduced in \cite{GOR-HOM-1}. 
In section 3   we get lower and upper  estimates of the growth rates corresponding  to an 
intermediate   fluid with constant viscosity. 
In section 4 we use this result  for a model with  $N$ intermediate layers with constant 
viscosities and we prove the flow instability.
In section 5 we get an intermediate  linear  viscosity profile  which can almost suppress 
the Saffman-Taylor instability.  We conclude in section 6.

\begin{center}

{\bf  2.  The three-layer Hele-Shaw model} 

\end{center}

The three-layer Hele-Shaw flow with variable  intermediate viscosity was first described in 
~\cite{GOR-HOM-1} and studied also in ~\cite{GOR-HOM-2}.  
 We recall here the  basic elements.  

A polymer solute with a variable concentration $c$ and variable viscosity $\nu$ is  injected  
with the positive velocity $U$ in a rectangular Hele-Shaw cell saturated with oil of viscosity 
 $ \nu_O$,  during a time interval $TI$.  
As in ~\cite{GOR-HOM-1},  adsorption, dispersion and diffusion of the solute in the equivalent 
porous medium are neglected. The  expression of the intermediate viscosity  $\nu$ in terms of 
$c$  is
\begin{equation}\label{ZT02}
 \nu(c)  = a_0 + a_1 c + a_2c^2 + .... 
\end{equation}
where $a_i$ are some coefficients which can depend on $x,y$ - see  ~\cite{GILJE},  ~\cite{FLORY}.
In the case  of a dilute  solute,  which is studied here, we have
$ \nu = a_0 + a_1 c $,  then $ \nu $ is invertible in  terms of $c$.
The continuity equation for the solute is $Dc/Dt=0$, then we have $D \nu / Dt=0$. That means 
\begin{equation}\label{ZT003}
\nu _t + u \nu_x  + v \nu_y = 0.
\end{equation}
}
At the end of the time interval $TI$,  a displacing fluid   with viscosity $\mu_W$ is injected in 
the porous medium,  with  the  same velocity $U$. 

We consider  incompressible fluids, then the amount of fluid between the two interfaces
cannot change, according to the  principle   of mass   conservation. Therefore an arbitrary (small) 
movement  of  one interface must induce a movement with the same velocity  of the other interface.

However, it is well known - see \cite{SAFF-TAY} - that interfaces  change over time and turn into 
fingers of fluid (or polymer solute).  We study the  evolution  of perturbations only  in a small 
time interval after $TI$ and  believe that the  initial  shape  of interfaces has not changed so  
much.
On this way we  obtain   an  intermediate  fluid layer, moving with the velocity $U$, where the 
viscosity is variable. Consider $u=U, v=0$, then from   \eqref{ZT003}   we get
$$  \nu = \nu(x-Ut) .                                                                          $$

An intermediate  polymer-solute with an exponentially- decreasing (from the front  interface) 
viscosity   $\nu(x-Ut)$  was used by  Mungan ~\cite{MUN} and  the instability  was almost  
suppressed. The displacements with variable   viscosity in Hele-Shaw cells   and   porous media  
are  studied in ~\cite{TALON},   ~\cite{LOGGIA}. 

It is possible  to inject several  polymer-solutes  with constant-concentrations  
$$c_1,c_2,..., c_N                                                                           $$ 
during the time intervals    
$$TI_1, TI_2, ..., TI_N.                                                                      $$ 
Then we obtain a steady flow of $N$ thin layers of immiscible fluids  with    {\it constant} 
viscosities  $\nu_i, \quad i=1,2,...,N$. This is the multi-layer model studied in  ~\cite{DAR-2},  
~\cite{DAR-DING},  \cite {D7} , \cite {D6}.

In this paper, the displacing fluid   is denoted with the lower index $_W$  and the  displaced 
one  with the  lower index  $_O$. 

\vspace{0.25cm}

Suppose the intermediate region  is the interval 
$$ Ut - Q  < x <  Ut,                             $$ 
moving with the  constant   velocity $U$ far upstream.   We have three incompressible   fluids with 
viscosities  $\nu_W$ (displacing fluid),  $ \nu $ (intermediate  layer) and  $ \nu_O$ (displaced 
fluid). The flow is governed by    the Darcy's  equations:  
$$  p_{x} = - \mu_d u; \quad  p_y = -\mu_d v;  
 \quad p_z=0;                                     $$
\begin{equation}\label{ZT004}
 u_{x} + v_y = 0;                                                           
\end{equation}
$$ \mu_d = \mu_W, \quad x < Ut-Q;                 $$
$$    \mu_d = \mu, \quad x \in ( Ut - Q,   Ut) ;  $$
$$     \mu_d = \mu_O, \quad x > Ut ;              $$
$$
\mu_W = 12\nu_W /b^2; \quad  \mu = 12 \nu /b^2;  $$
\begin{equation}\label{DIM-VISCO}
\mu_O =  12 \nu_O/b^2.                                                                   
\end{equation}
	
\vspace{0.25cm}	

 The  basic velocity and interfaces are 
$$ u=U, \,\, v=0; \quad  \quad  x = Ut-Q, \,\, x =Ut. $$ 
On the interfaces we consider the  Laplace's law: the 
pressure jump is given by the surface tension multiplied with 
the interfaces curvature and the  component $u$ of the velocity 
is continuous.  Moreover,  the interface is a material one. The 
basic interfaces are straight lines,  then the basic
 pressure $P$ is  continuous (but his gradient is not) and          
 \begin{equation}\label{BASIC-PRESS}
P_{x}= -  \mu_d U, \quad P_y=0.
\end{equation}
We use the equation  \eqref{ZT003}, then the basic (unknown) 
viscosity $\mu$ in the middle layer verifies  the equation
\begin{equation}\label{ZT004A}
 \mu_t + U \mu_{x} = 0.
\end{equation}
We introduce the moving reference frame 
\begin{equation}\label{MR}
{\overline x}=x - Ut, \quad    \tau = t.                
\end{equation}
The equation  \eqref{ZT004A}  leads to $\mu_{\tau}=0$, then  
$\mu = \mu ({\overline x}) $. The middle region in the  moving  
reference frame is the segment  $ -Q <{\overline x} < 0 $. However,  we 
still use  the notation  $x, \,\, t$ instead of 
${\overline x}, \tau$.

\vspace{0.25cm}

The perturbations   of the basic velocity, pressure and viscosity are 
denoted   by $u', v', p', \mu' $.  We  insert the perturbations  in the 
equations    \eqref{ZT004}, \eqref{ZT004A}. As in  ~\cite{GOR-HOM-1},  
we obtain the  linear stability  equations which governs the small
perturbations:
$$
 p'_x = -\mu u' - \mu' U, \quad p'_y = -  \mu v',                $$
\begin{equation}\label{ZT005A}
 u'_x + v'_y = 0,
\end{equation}
\begin{equation}\label{ZT006A}
 \mu'_t + u' \mu_x = 0. 
\end{equation}
A Fourier decomposition for the perturbation $u'$  is used:
$$
 u'(x,y,t) =                                                       $$
\begin{equation}\label{FOURIER-U}
f(x)  [   \cos(ky) + \sin (ky)] e^{\sigma t}, \,\, k \geq 0,  
\end{equation}
where   $f(x)$ is the amplitude, $\sigma $ is the growth constant  and  
$k$ are the wave numbers.The dimension 
of $f$ is {\it (space)/(time)}.

\vspace{0.25cm}

As the  velocity along the axis $Ox$ is continuous, the amplitude 
$f(x)$ is  continuous. From   $\eqref{ZT005A}_3$,  $\eqref{ZT005A}_2$, 
\eqref{ZT006A},  \eqref{FOURIER-U}  we get  the Fourier decompositions
for the perturbations $v', p', \mu'$:
$$ v' = ( 1/k) f_x 
[-  \sin(ky) + \cos(ky)] e^{\sigma t},                                      $$
$$ p' = (\mu / k^2) f_x 
[-  \cos(ky) - \sin(ky)] e^{\sigma t},                                      $$                               
\begin{equation}\label{ZT007}
 \mu' = (-1/ \sigma) \mu_x f
[  \cos(ky) +  \sin (ky)]e^{\sigma t}.           
\end{equation} 
The cross derivation of the relations $\eqref{ZT005A}_1, 
\eqref{ZT005A}_2$    leads us to
$$ \mu u'_y + \mu'_y U =  \mu_x v' + \mu v'_{x} .              $$
Then from  $\eqref{FOURIER-U}, \eqref{ZT007}_1,  \eqref{ZT007}_3$ 
we get the  equation which  governs   the amplitude  $f$:
\begin{equation}\label{ZT008}
 -(\mu f_x)_x +  k^2 \mu f = \frac{1}{\sigma} U k^2 f \mu_x,  
 \quad \forall x \notin \{-Q,0 \}.
\end{equation}
The viscosity is constant outside the intermediate region, then  
\eqref{ZT008}  becomes 
$$ -f_{xx} + k^2 f = 0, \quad x \notin (-Q,0)    .              $$
The perturbations must decay to zero 
in   the far field and $f$ is continuous  and  we have
$$
f(x) = f(-Q) e^{ k(x+Q) }, \,\, \forall x \leq -Q;              $$
\begin{equation}\label{FAR-FIELD}
f(x) = f(0) e^{ -kx }, \,\, \forall x   \geq 0.
\end{equation}
 
\vspace{0.25cm}

We now describe the Laplace law 
in  a  point $a$ where a  a viscosity jump exists.
The amplitude $f$ is continuous in $a$ but we  have a jump of 
$f_x$. 
The perturbed interface near $a$ is denoted by $\eta(a,y,t)$. 
In the first approximation we have  $\eta_t = u$, therefore   
$$
\eta(a,y,t) =                                                   $$
\begin{equation}\label{INTER001}
(1/ \sigma)  f(a) 
[  \cos(k y) +  \sin(ky) ] e^{\sigma t }.
\end{equation}  
We search  for the right and left limit values  of the pressure in 
the point   $a$, denoted  by $p^+(a), \quad p^-(a)$. 
For this we use  the basic pressure $P$   in the point  $a$, the 
Taylor first order  expansion  of $P$ near $a$ and the expression   
$\eqref{ZT007}_2$ of $p'$ in $a$. From \eqref{BASIC-PRESS}  it  
follows  $P_x^{+,-}(a)= -\mu^{+,-}(a)U $   then  we get 
$$
p^+(a) = P^+(a) + P^+_x(a) \eta + p'^+(a) =   P^+(a)       $$
\begin{equation}\label{INTER002}
- \mu^+(a)  \{ \frac{U f(a)}{\sigma} +\frac{ f_x^+(a)}{k^2} \}
[ \cos(k y) +  \sin(ky) ]e^{\sigma t},      
\end{equation}              
$$
p^-(a) = P^-(a) + P^-_x(a) \eta + p'^-(a) =   P^-(a)        $$
\begin{equation}\label{INTER003}
- \mu^-(a)  \{ \frac{U f(a)}{\sigma} +\frac{ f_x^-(a)}{k^2} \}
[ \cos(k y) +  \sin(ky) ]e^{\sigma t},       
\end{equation}               
The  Laplace's law is
\begin{equation}\label{LAPLACE001}
  p^+(a) - p^-(a) = T(a) \eta_{yy},
\end{equation}
where $T(a)$ is the surface tension acting in the point $a$ 
and  $\eta_{yy}$  is the approximate value  of the  curvature 
of  the perturbed  interface. As $P^-(a)=P^+(a)$,  from the 
equations \eqref{INTER002} - \eqref{LAPLACE001} we get  the 
relationship between    $f_x^-(a)$, $ f_x^+(a)$ and $\sigma$:
$$
- \mu^+(a)[\frac{Uf(a)}{\sigma}+  \frac{f_x^+(a)}{k^2}] +     $$
$$  \mu^-(a)[\frac{Uf(a)}{\sigma}+\frac{f_x^-(a)}{k^2}] =     $$   
\begin{equation}\label{LAPLACE002}
-  \frac{T(a)}{\sigma}f(a)k^2.
\end{equation}

\vspace{0.25cm}

 The growth constant  for three-layer case is obtained  as follows. 
We  multiply with $f$ in the amplitude equation  \eqref{ZT008}, we 
integrate on $(-Q,0)$ and obtain  
$$  - \int_{-Q}^0 (\mu f_x f)_x +  
\int_{-Q}^0 \mu f_x^2 +     k^2 \int_{-Q}^0 \mu f^2 =      $$
$$ \frac{k^2U}{\sigma} \int_{-Q}^0 \mu_x f^2,              $$
therefore
$$
 \mu^+(-Q)f_x^+(-Q)f(-Q) -  \mu^-(0)f_x^-(0)f(0) +         $$
\begin{equation}\label{PRE-SIGMA}
 \int_{-Q}^0 \mu f_x^2 +     k^2 \int_{-Q}^0 \mu f^2 =   
\frac{k^2U}{\sigma} \int_{-Q}^0 \mu_x f^2.                   
\end{equation}

From the relations \eqref{FAR-FIELD}  we have   
$$
f_x^-(-Q) =  k f_1  \,\, f_x^+(0)=-k f_{0}, $$  
\begin{equation}\label{PRE-SIGMA-B}         
  f_1 = f(-Q), \quad f_{0} =f(0)
\end{equation}

Recall  $ \mu^-(-Q)= \mu_W, \,\,
\mu^+(0)= \mu_O$, then  from  \eqref{LAPLACE002}, \eqref{PRE-SIGMA},  
\eqref{PRE-SIGMA-B} it  follows
$$
 \sigma = \frac{ S_0 f_0^2 + S_1 f_1^2 + 
k^2 U \int_{-Q}^0 \mu_x f^2}
{\mu_Ok f^2_0 +  \mu_W kf^2_1 + I},                      $$
$$   S_0 =k^2U(\mu^+ - \mu^-)_0 - k^4T_0,                   $$
$$   S_1 = k^2U(\mu^+ - \mu^-)_{-Q} - k^4 T_1,        $$
\begin{equation}\label{SIGMA001}
   I =  \int_{-Q}^0 [ \mu f_x^2 + k^2 \mu f^2 ],           
\end{equation}
where $T_0, T_1$ are the surface tensions in $x=0, x=-Q$.
The viscosity jumps in the above expression are
$$(\mu^+ - \mu^-)_0= \mu_O-\mu^-(0),                          $$
$$ (\mu^+ - \mu^-)_{-Q}= \mu^+(-Q) -\mu_W.                    $$
We have to find   the  basic viscosity $\mu$  which  minimizes  
the  growth constant  $\sigma$.

\vspace{0.25cm}

{\it Remark 1}. We suppose that  in  $x=a$ exist:

i)  a viscosity jump $(\mu_O - \mu_W)$; \hspace{1cm} 
ii) a surface tension $T(a)$.           \hspace{1cm} Then
$$-f_{xx}+k^2f=0, \,\, x \neq a; \quad  
 f(x) = f(a) e^{ k(x-a) },  \, x \leq a; \,
  f(x) = f(a) e^{ -k(x-a) }, \, x \geq a                          $$
and  from  \eqref{LAPLACE002}   we recover the SaffmanTaylor  formula
\begin{equation}\label{SIGMA_ST01} 
\sigma_{ST} = \frac{ k U(\mu_O - \mu_W) - T(a) k^3}{\mu_O + \mu_W}.
\end{equation}
If $\mu_O > \mu_W$,  then $\sigma_{ST}>0$ in the range
$$  k^2 <  U(\mu_O - \mu_W)/T(a)                               $$
and the flow is unstable.  We  also  have 
\begin{equation}\label{SIGMA_ST02}
T(a) = 0 \quad  \Rightarrow 
  \sigma_{ST} =  k U \frac{(\mu_O - \mu_W)}{\mu_O + \mu_W}. 
\end{equation}

$ \hfill \square $

\begin{center}

{\bf  3.  The three-layer case   with constant  viscosities and zero surface tensions }

\end{center}

We consider a constant intermediate viscosity $\mu_1 \in  (\mu_W,  \mu_O) $.                                                   
The corresponding  growth rate is denoted by $\sigma_1$.  
The formula \eqref{SIGMA001} with $T_0=T_{-Q}=0$ and notations \eqref{PRE-SIGMA-B}  
becomes 
$$
 \sigma_1 =k^2{\color{red} U}  \frac{(\mu_1-\mu_W)f^2_1 + (\mu_O-\mu_1)f^2_0}
{\mu_W kf^2_1 + \mu_Ok f^2_0 + I_1},                                          $$
\begin{equation}\label{SIGMA001-CONST} 
  I_1 =  \mu_1 \int_{-Q}^0 ( f_x^2 + k^2 f^2 ).                         
\end{equation}
In this section we prove that $\sigma_1 \rightarrow \infty $ for large  $k$. 

{\it Lemma 1. } If 
$$
-f_{xx} + k^2 f = 0, \quad  \forall x \in  (a,c),       $$
\begin{equation}\label{I001} 
 I(a,c) =  \int_{a}^{c} ( f_x^2+ k^2 f^2 ),                                 
\end{equation}
then 
$$
\quad k \frac{ e^{k(c-a)} - 1}{e^{k(c-a)}+1}[ f^2(a)+ f^2(c) ] 
\leq I(a,c) \leq                                        $$
\begin{equation} \label{INEQ003}
 k \frac{ e^{k(c-a)} + 1}{e^{k(c-a)}-1}[ f^2(a)+ f^2(c) ].
\end{equation}

{ \it Proof.}  The solution of the equation $\eqref{I001}_1$ is given by  
$ f(x) = A e^{kx} + B e^{-kx}$,  
where $A,B$ are constant with respect to  $x$. We multiply the equation 
$\eqref{I001}_1$  with $f$ and get 
$$ f_x^2+ k^2 f^2 = (f_xf)_x,                                          $$ 
then from $\eqref{I001}_2$  it 
follows 
$$ I(a,c)  = (f_xf)(c) -(f_xf)(a) =                                    $$
$$  k(A e^{kc} -  B e^{-kc})(A e^{kc} +  B e^{-kc})                    $$         
$$ - k(A e^{ka} -  B e^{-ka})(A e^{ka} +  B e^{-ka}) =                 $$
$$ k[ A^2 e^{2kc} - B^2 e^{-2kc} - A^2 e^{2ka} + B^2 e^{-2ka}] =       $$
\begin{equation} \label{IN004}
 k \frac{ e^{2kc}-e^{2ka}}{e^{2k(a+c)}}[ A^2 e^{2k(a+c)} + B^2].
\end{equation}                       
Therefore   we have
$$
I(a,c)  = k \frac{ e^{2kc}-e^{2ka}}{e^{2k(a+c)}} D,                     $$
\begin{equation} \label{IN004A}
D=  A^2 e^{2k(a+c)} + B^2.
\end{equation}
We use the notation 
$$f_0 = f(c), \quad f_1 = f(a)                                         $$ 
and get
$$ A = \frac{f_0 e^{kc}- f_1 e^{ka}}{e^{2kc}- e^{2ka}},                $$        
$$ B= \frac{- f_0 e^{ka}+ f_1 e^{kc}}{e^{2kc}- e^{2ka}}e^{k(a+c)},     $$ 
therefore the relation  $\eqref{IN004A}_2$  gives us
$$
D  = C \frac{e^{2k(a+c)}}{(e^{2kc}- e^{2ka})^2},                       $$  
\begin{equation} \label{IN004A1} 
C= (f_0^2 + f_1^2) (e^{2kc}+e^{2ka})- 4 f_0f_1 e^{k(a+c)}.  
\end{equation}

In the expression $\eqref{IN004A1}_1$ we add and subtract 
$2(f_0^2+f_1^2)e^{k(a+c)}$, then  $C$ becomes 
$$ C = (f_0^2 + f_1^2) (e^{kc}-e^{ka})^2 +                             $$              
$$ 2(f_0^2+f_1^2- 2f_0f_1) e^{k(a+c)}.                                 $$

We have   $(f_0^2+f_1^2- 2f_0f_1)   \geq 0$, then the  formulas  $\eqref{IN004A1}$   
give us 
\begin{equation} \label{IN004A2} 
D \geq {{} \frac{e^{2k(a+c)}}{(e^{2kc}- e^{2ka})^2} } 
(e^{kc}-e^{ka})^2 (f_0^2 + f_1^2) .                                       
\end{equation}

On the other hand we have
$$ - 4 f_0f_1 e^{k(a+c)} \leq 2(f_0^2 + f_1^2)e^{k(a+c)}             $$ 
and by using this inequality in $\eqref{IN004A1}_2$ we get
\begin{equation} \label{IN004A3}
 D  \leq {{} \frac{e^{2k(a+c)}}{(e^{2kc}- e^{2ka})^2} } 
(e^{kc}+ e^{ka})^2 (f_0^2 + f_1^2).                                        
\end{equation}
The estimates  \eqref{IN004A2},  \eqref{IN004A3} and the formulas 
 \eqref{IN004A} - \eqref{IN004A1} are giving us  the inequalities  
\eqref{INEQ003}. 

$\hfill \square $

We use  {\it Lemma 1}  with $a=-Q, c=0 $,  then for {\it large 
 enough} $k$ we obtain 
\begin{equation} \label{L-00}
 I(- Q,0) \approx k [f^2_1 + f^2_0].                               
\end{equation}
Let  $m, x, n, y, M, N  >0$.  We have the inequalities
$$
min\{ \frac{M}{m}, \frac{N}{n} \}  \leq \frac{ Mx+Ny}{mx+ny} \leq       $$  
\begin{equation} \label{L-01}
max\{  \frac{M}{m}, \frac{N}{n}  \}.
\end{equation}
From \eqref{SIGMA001-CONST},  \eqref{INEQ003}, \eqref{L-00}, \eqref{L-01} 
  with 
$$m=(\mu_1+\mu_W), \, n =(\mu_O+\mu_1), \,\,\,  x= f^2_1,\, y= f^2_0, $$
$$M=(\mu_1-\mu_W), \,\,  N= \mu_O-\mu_1 ,                                $$ 
 we get  a formula somewhat similar to \eqref{SIGMA_ST02}: 
\begin{equation} \label{L-02}
 \sigma_1 \geq k {\color{red}U} min \{ \frac{ \mu_1-\mu_w}{\mu_1+\mu_W},
\,\, \frac{ \mu_O-\mu_1}{\mu_O+\mu_1}  \} .                                 
\end{equation}
Therefore   we obtain the following

{\it Proposition 1. }
The growth rate corresponding to the  constant  intermediate viscosity $\mu_1$
is unbounded with respect to  the wave numbers $k$  of perturbations.

$\hfill \square $


\begin{center}

 {\bf 4. The  $N$-layers Hele-Shaw model with constant viscosities} 

\end{center}

We consider   $N>1$ and  we divide the middle region in $N$ small intervals 
(layers) $(x_{i+1},x_i)$  of length $(Q/N)$,   where  
\begin{equation}\label{PUNCTELE}
 x_i = -i Q/N, \quad   i=0,1,...N,                                    
\end{equation}
are the interfaces between the layers.    All surface tensions  on 
interfaces are zero.  On each small interval, for $i=1,2,...,N$  we 
have   the constant viscosities 
\begin{equation}\label{STRATELE}
\mu_i = \mu_O - i(\mu_O-\mu_W)/(N+1),                                        
\end{equation}                      
and  the  amplitude equations 
\begin{equation}\label{STRATELE-2}
 -\mu_i f_{xx} + \mu_ik^2f =0.                                
\end{equation} 
The corresponding growth  constants are denotd by  $\sigma_N$.
In this section we prove that  
$$\sigma_N \rightarrow {\infty} \quad \mbox{ for large } 
\quad k.                                                    $$

We multiply with $f$  in all  equations \eqref{STRATELE-2} and use 
the boundary  conditions \eqref{LAPLACE002} in each point $a= x_i$ 
where a viscosity jump exists. We integrate on $(-Q, 0)$. The 
 method used in section 2 gives us  the  following formula of the 
growth constant denoted  by $\sigma_N$  (see also the corresponding   
expression in ~\cite{DAR-2}  with all surface tensions zero):
$$
 \sigma_N = \frac{\sum_{i=0}^{i=N} k^2U(\mu^+ - \mu^-)_i f_i^2}
{ k\mu_W f_N^2 +  k\mu_O f_0^2 +  \sum_{i=1}^{i=N} I_i},           $$
$$
I_i = \int_{x_i}^{x_{(i-1)}}  \mu_i (f_x^2 + k^2 f^2 ), \quad
 f_i=f(x_i),                                                       $$
$$
(\mu^+ - \mu^-)_i= \mu_i-\mu_{i+1}=                                $$
\begin{equation}\label{SIG00N}
(\mu_O-\mu_W)/{{} (N+1).}
\end{equation}

{\it Proposition 2 }. For large  enough $k$ we have
\begin{equation}\label{SIGMA00NA3A}  
\sigma_N  \geq  k  U  
\frac{(\mu_O-\mu_W)} {  (2N+1)\mu_O + \mu_W }.
\end{equation}

{\it Proof}.  We recall the notations \eqref{PUNCTELE}, \eqref{STRATELE}, 
\eqref{SIG00N}    and consider  
$$ a_i=x_{i+1}, \quad  c_i=x_i                                       $$  
then
$$f_{i+1}= f(a_i), \quad f_i= f(c_i).                                $$ 
We have  $c_i-a_i = Q/N$, then from { \it Lemma 1 } we get
$$
\int_{a_i}^{c_i}  (f_x^2 + k^2 f^2 ) \leq  
k \Theta(k)   [ f_{i+1}^2+ f_i^2 ],                                  $$
\begin{equation} \label{SIGMA00N2}
\Theta(k) = \frac{ e^{kQ/N} + 1}{e^{kQ/N}- 1}. 
\end{equation}                  

The next inequalities can be easily verified:
$$
 A_i, B_i, x_i >0 ,  \quad i=0,1,...,N     \Rightarrow                $$  
\begin{equation}\label{INEQMIN}
min \frac{A_i}{B_i}  \leq \frac{ \sum_{i=0}^{i=N} A_ix_i}
     { \sum_{i=0}^{i=N} B_ix_i} 
\leq max  \frac{A_i}{B_i}.                           
\end{equation}
From $\eqref{SIG00N}_1$, \eqref{SIGMA00N2}, \eqref{INEQMIN}  we obtain 
$$
\sigma_N \geq k U \frac{\mu_O-\mu_W}{ (N+1)} 
 Min \{ G_1, G_i, G_N \},                                              $$
$$  G_1=  \frac{1}{\mu_O + \mu_1 \Theta(k)},                           $$ 
$$  G_i=  \frac{1}{\Theta(k)[ \mu_{i+1} + \mu_i]},                     $$
\begin{equation}\label{SIGMA00NA2}
  G_N = \frac{1}{\mu_W + \mu_N \Theta(k)}.                          
\end{equation}
The equation $\eqref{SIGMA00N2}_2$ for  large $k$ is giving   
$\Theta(k) \approx 1$,  then  
$$  Min \{ G_1, G_i, G_N \} =  \frac{1}{\mu_O + \mu_1},                $$
and the above   estimate $\eqref{SIGMA00NA2}_1$    leads to
\begin{equation}\label{SIGMA00NA2A} 
\sigma_N \geq k U \frac{\mu_O-\mu_W}{{{} (N+1)}}  \times  
\frac{1}{\mu_O + \mu_1}.                                             
\end{equation}
We obtain $\mu_1$ from   \eqref{STRATELE}  and the inequality  \eqref{SIGMA00NA3A}
follows from  the estimate    \eqref{SIGMA00NA2A}. 
   
$ \hfill \square $

If   all involved surface tensions are not zero 
and verify some conditions, then  the growth constants corresponding to the $N$
-layer model with the intermediate viscosities  \eqref{STRATELE} can be arbitrary 
small (positive) if $N$ is large enough - see ~\cite{DAR-2}, ~\cite{DAR-DING}.

\vspace{0.25cm}

{\it Remark 2}. We consider the case when  the intervals $(x_{i+1},x_i)$
are not equals  and  $\mu_i$  are verifying 
$ \mu_O >\mu_1 > \mu_2 > ... \mu_{N} >\mu_W.                             $  
The corresponding growth constant is denoted by $\sigma_{Nneq}$.
 {\it Lemma 1}, \eqref{SIG00N}  and   \eqref{INEQMIN} lead us to the following
estimate
\begin{equation}\label{SIG-NEC}
  \sigma_{Nneq} \geq k U  Min \{ H_0, H_i, H_1 \},
\end{equation}
$$ H_0 = \frac{\mu_O-\mu_1}{\mu_O + \mu_1 \Gamma_1(k)},                    $$
$$ H_i = \frac{(\mu_{i} - \mu_{i+1})}
{\Gamma_{i}(k) \mu_{i} + \Gamma_{i+1}(k)\mu_{i+1}},                        $$
\begin{equation}\label{SIGMA00NA32}
H_N = \frac{\mu_{N} - \mu_W}{\mu_N \Gamma_N(k) + \mu_W },
\end{equation}
$$
\Gamma_i(k)= 
\frac{ e^{k(x_{i-1}-x_i)} + 1}{e^{k(x_{i-1}-x_i)}-1}.                        $$
$\hfill \square$

 $$    \mbox{ {\bf  5. The  three-layers model with linear  intermediate viscosity}   }    $$

We consider the formula \eqref{SIGMA001} with  $T_0 =T_1=0 $ and the  
viscosity profiles plotted in the Figures 1 a) - d) below, therefore 
\begin{equation}\label{SIGMA_ST01A} 
(\mu^+ - \mu^-)_0 \leq 0,  \,\,  (\mu^+ - \mu^-)_{-Q} \leq 0.  
\end{equation}

We prove that the corresponding growth constants (denoted by $\sigma_L)$ 
are bounded with respect to  $k$, even if  both  surface tensions are zero. 
In the formula  \eqref{SIGMA001}, we neglect the  viscosity jumps in the 
numerator,   the positive terms
$\mu_Ok f^2_0, \,\,  \mu_W kf^2_{-Q}, \,\, \int_{-Q}^0  \mu f_x^2    $  
in the  denominator  and  obtain the upper estimate   below:
\begin{equation}\label{SIGMA002}
\sigma_L \leq  \frac{ k^2 U \int_{-Q}^0 \mu_x f^2} 
                { {{} \int_{-Q}^0  k^2 \mu f^2 } }  =
U \frac{\int_{-Q}^0 \mu_x f^2} { \int_{-Q}^0  \mu f^2 }.
\end{equation}

\vspace{0.5cm}

\begin{figure*} [h]
\centering
\includegraphics[scale=0.5]{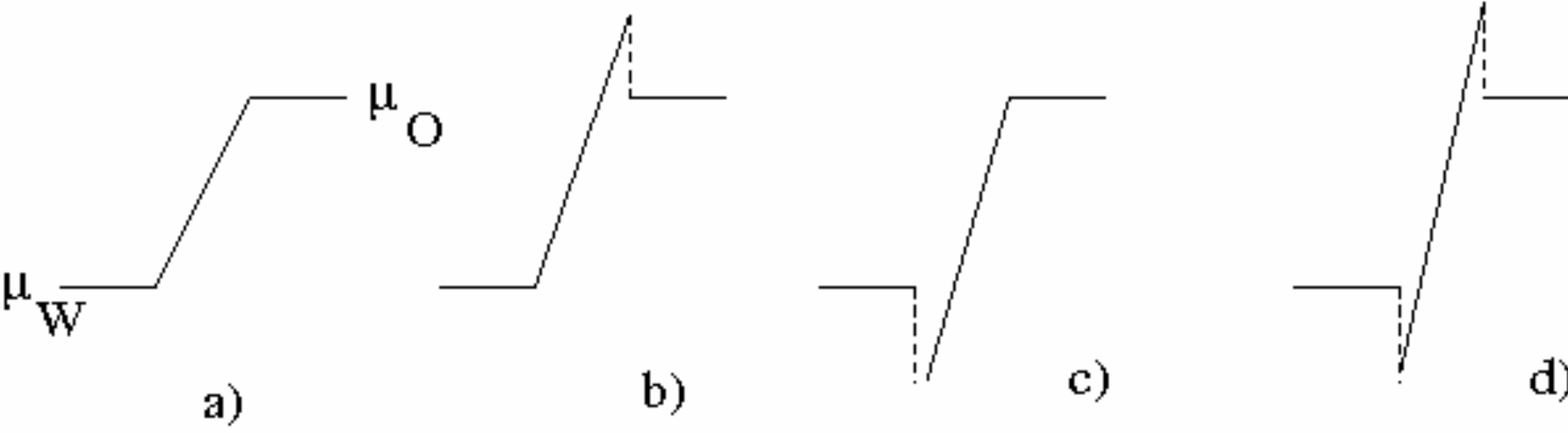}
\vspace{-12cm}
\caption{a) continuous linear viscosity between displacing fluid  
and oil;  b), c), d): discontinuous linear viscosities with { \it 
negative}  jumps in $x=0$ or (and)  $x=-Q$.  }
\end{figure*}


\vspace{0.25cm}

Let  $\mu_{min}$ be  the smallest            value of $\mu$  in the 
intermediate region, which can be less than $\mu_W$, as in Figures 1 
c) - d).
We have $\mu_x >0 $, and from  \eqref{SIGMA002} we get
\begin{equation}\label{SIGMA003}
\sigma_L \leq U \frac{ Max_x (\mu_x ) } {  \mu_{min}  }.
\end{equation}
The above upper bound is not depending on the  maximum value of the 
viscosity,  but only on the {\it  maximum value } of his derivative 
and on  $\mu_{min}$.

\vspace{0.5cm}

{\it Remark 3.} The total (dimensional)  amount $TA$ of liquid introduced 
in intermediate region  is given by (see \cite{GOR-HOM-1}) 
\begin{equation}\label{TA-1}
TA = \int_{-Q}^0 \mu(x) dx.
\end{equation}
We prove that $TA$ is the same for  the viscosity profile given in Figure 1 
a)  and  for the  $N$ layer flow described  by the formulas  \eqref{PUNCTELE} - 
\eqref{STRATELE}. For the linear profile the Figure 1a)   we have
$$\int_{-Q}^0 \mu(x) dx =  (\mu_W  +  \mu_O)Q/2                            $$
and for  the $N$ layer viscosity profile \eqref{PUNCTELE} -  \eqref{STRATELE}
we obtain the same result:
$$ \int_{-Q}^0 \mu(x) dx = Q\mu_O -
\sum_{i=N}^{i=1}  i( \frac{\mu_O-\mu_W}{N+1} ) \frac{Q}{N}.                $$

{ \it Remark 4.}
The { \it linear}  continuous viscosity profile  plotted in the Figure 1a) and 
the estimate  \eqref{SIGMA003} give us
\begin{equation}\label{SIGMA004}
 \sigma_{LC} \leq U \frac{\mu_O - \mu_W}{Q \mu_W}.
\end{equation}  
Therefore we get an arbitrary small  positive growth constant if  $Q$
 is large enough, even if both surface tensions in $x=-Q, \quad x=0$ 
are zero.

$ \hfill \square $

We mention here that on the page 3 of ~\cite{GEOLOGIC} 
is considered a linear viscosity profile in a porous medium.

\begin{center}

 {\bf  6. Conclusions}

\end{center}

The interface between two Newtonian immiscible fluids   in a rectangular Hele-Shaw cell is 
unstable  when the  displacing fluid is less viscous. If the surface tension on
the interface is zero, then the  Saffman-Taylor growth  constant of the linear 
perturbations is boundless with respect to  the wave numbers  $k$ - see the formula 
\eqref{SIGMA_ST02}.

 An intermediate fluid  with  a variable  viscosity  between the 
displacing fluid  and oil can minimize  the  Saffman-Taylor instability when the 
surface tensions are different from zero
 - see  the papers  \cite{GILJE}, ~\cite{GOR-HOM-1},  ~\cite{GOR-HOM-2},  
~\cite{SHAH},  ~\cite{SLOBOD},    ~\cite{UZOIGWE}.

The multi-layer Hele-Shaw model, consisting  of $N$ intermediate fluids with constant 
viscosities was studied  in 
~\cite{DAR-2},  ~\cite{DAR-DING},  \cite {D7} , \cite {D6} and upper bounds of the 
growth rates   were obtained. 
If  all surface tensions   are {\it different from  zero} and verify some conditions,  
an arbitrary small (positive) upper bound of the growth rates can be obtained, if $N$ 
is large enough. This  model is useless when  all surface tensions on the interfaces 
are   zero.

In this paper we study the Hele-Shaw displacement in rectangular  cells, when all surface 
tensions on the interfaces are zero. 

We point  out  a significant difference between the displacement with constant intermediate 
viscosities and the  displacement with a single variable intermediate viscosity. In the first 
case, if the viscosity-jumps are positive in the flow direction, then the displacement 
process is unstable - see   {\it Proposition 2}. In the  second case  we can almost suppress 
the  Saffman-Taylor instability. 

We get   lower bounds of the growth rates in the
three-layer case with constant intermediate viscosity - see {\it Lemma 1} in section 3.
We  use this result for the case of $N$ intermediate constant-viscosity layers  and get
the lower bounds \eqref{SIGMA00NA3A} and  \eqref{SIG-NEC}. Therefore  the growth 
rates are unbounded with respect to the wave numbers of perturbations, as in  the 
Saffman-Taylor  case without surface tension.

In section 5 we study the three-layer case without surface tensions. An intermediate  fluid 
with a linear increasing   viscosity gives us arbitrary small (positive) growth constants if 
the middle region   is large enough - see the formula \eqref{SIGMA004}.

The total amount of intermediate liquid for the $N$ layer flow given by  \eqref{PUNCTELE},  
\eqref{STRATELE}  and  for the variable linear viscosity-profile given in Figure  1a) is the 
same - see    {\it Remark 3}.

Our main conclusion is following.  When  all surface tensions are zero, the  best strategy to minimize 
the Saffman-Taylor  instability  is to use  an intermediate  liquid  with a suitable {\it variable} 
viscosity. On this way we can almost suppress the  instability.

\end{document}